\documentclass[sn-basic]{sn-jnl}
\usepackage[T1]{fontenc}
\usepackage[utf8]{inputenc}
\usepackage{natbib,aas_macros}

\jyear{2022}%

\theoremstyle{thmstyleone}%
\theoremstyle{thmstyletwo}%

\theoremstyle{thmstylethree}%

\begin{document}

\title[Accretion disks, quasars and cosmology: 
meandering towards understanding]{Accretion disks, quasars and cosmology: 
meandering towards understanding}


\author*[1]{\fnm{Bo\. zena} \sur{Czerny}}\email{bcz@cft.edu.pl}

\author[2]{\fnm{Shulei} \sur{Cao}}\email{shulei@phys.ksu.edu}

\author[1]{\fnm{Vikram Kumar} \sur{Jaiswal}}\email{vkj005@gmail.com}

\author[3]{\fnm{Vladim\'{\i}r} \sur{Karas}}\email{vladimir.karas@asu.cas.cz}

\author[4]{\fnm{Narayan} \sur{Khadka}}\email{nkhadka@phys.ksu.edu}

\author[5]{\fnm{Mary Loli} \sur{Mart\'inez-Aldama}}\email{mmartinez@das.uchile.cl}

\author[1]{\fnm{Mohammad Hassan} \sur{Naddaf}}\email{naddaf@cft.edu.pl}

\author[6]{\fnm{Swayamtrupta} \sur{Panda}}\email{spanda@lna.br}

\author[7]{\fnm{Francisco} \sur{Pozo Nu\~nez}}\email{francisco.pozonunez@h-its.org}

\author[1]{\fnm{Raj} \sur{Prince}}\email{raj@cft.edu.pl}

\author[2] {\fnm{Bharat} \sur{Ratra}}\email{ratra@phys.ksu.edu}

\author[8,1]{\fnm{Marzena} \sur{Sniegowska}}\email{marzena.sniegowska@gmail.com}

\author[9]{\fnm{Zhefu} \sur{Yu}} \email{yu.2231@buckeyemail.osu.edu}

\author[10]{\fnm{Michal} \sur{Zaja\v{c}ek}}\email{zajacek@mail.muni.cz}

\affil*[1]{\orgdiv{Center for Theoretical Physics}, \orgname{Polish Academy of Sciences}, \orgaddress{\street{Al. Lotnik\' ow 32/46}, \city{Warsaw}, \postcode{02-668}, \country{Poland}}}

\affil[2]{\orgdiv{Department of Physics}, \orgname{Kansas State University}, \orgaddress{\street{116 Cardwell Hall}, \city{Manhattan}, \postcode{KS 66506}, \country{USA}}}

\affil[3]{\orgdiv{Astronomical Institute}, \orgname{Czech Academy of Sciences}, \orgaddress{\street{Bo\v{c}n\'{\i}}~II~1401}, \city{Prague}, \postcode{CZ-14100},  \country{Czech Republic}}

\affil[4]{\orgdiv{Department of Physics}, \orgname{Bellarmine University}, \orgaddress{\street{2001 Newburg Rd}, \city{Louisville}, \postcode{KY 40205}, \country{USA}}}

\affil[5]{\orgdiv{Departamento de Astronom\'ia}, \orgname{Universidad de Chile}, \orgaddress{\street{Camino del Observatorio 1515}, \city{Santiago}, \postcode{Casilla 36-D Correo Central}, \country{Chile}}}

\affil[6]{\orgdiv{Laborat\'orio Nacional de Astrof\'isica}, \orgname{MCTI}, \orgaddress{\street{R. dos Estados Unidos 154}, \city{Itajub\'a}, \postcode{37504-364}, \country{Brazil}}}

\affil[7]{\orgdiv{Astroinformatics}, \orgname{Heidelberg Institute for Theoretical Studies}, \orgaddress{\street{Schloss-Wolfsbrunnenweg 35}, \city{Heidelberg}, \postcode{69118}, \country{Germany}}}

\affil[8]{\orgdiv{Copernicus Astronomical Center}, \orgname{Polish Academy of Sciences}, \orgaddress{\street{Bartycka 18}, \city{Warsaw}, \postcode{00-716}, \country{Poland}}}

\affil[9]{\orgdiv{Department of Astronomy}, \orgname{Ohio  State University}, \orgaddress{\street{Ohio 43210}, \city{Columbus}, \postcode{Ohio 43210}, \state{Ohio}, \country{USA}}}

\affil[10]{\orgdiv{Department of Theoretical Physics and Astrophysics}, \orgname{Faculty of Science, Masaryk University}, \orgaddress{\street{Kotl{\'a}\v{r}sk{\'a} 2}, \city{Brno}, \postcode{611 37}, \country{Czech Republic}}}


\abstract{As Setti and Woltjer noted back in 1973, one can use quasars to construct the Hubble diagram; however, the actual application of the idea was not that straightforward. It took years to implement the proposition successfully. Most ways to employ quasars for cosmology now require an advanced understanding of their structure, step by step. We briefly review this progress, with unavoidable personal biases, and concentrate on bright unobscured sources. We will mention the problem of the gas flow character close to the innermost stable circular orbit near the black hole, as discussed five decades ago. This problem later led to the development of the slim disk scenario and is recently revived in the context of Magnetically Arrested Disks (MAD) and Standard and Normal Evolution (SANE) models. We also discuss the hot or warm corona issue, which is still under debate and complicates the analysis of X-ray reflection. We present the scenario of the formation of the low ionization part of the Broad Line Region as a failed wind powered by radiation pressure acting on dust (Failed Radiatively Driven Dusty Outflow -- FRADO). Next, we examine the cosmological constraints currently achievable with quasars, primarily concentrating on light echo methods (continuum time delays and spectral-line time delays to the continuum) that are (or should be) incorporating the progress mentioned above. Finally, we briefly discuss prospects in this lively subject area.}




\maketitle

\section{Introduction}\label{sec1}

The purpose of this article is to describe many steps in our understanding of how Active Galaxies operate, to finally open up the possibility to use them as exploratory tools in cosmology. 

This discussion is timely as one recognizes specific cracks reported in the current picture of the Universe as represented by the standard $\Lambda$CDM model. The most spectacular of these cracks is the tension in the Hubble constant. Especially between the values derived from some local measurements of the Universe expansion (of the order of 74 km s$^{-1}$ Mpc$^{-1}$, see e.g. \citealt[][]{wong2020} and \citealt{riess2022}), and the value expected from the analysis of the Cosmic Microwave Background (CMB) in the standard $\Lambda$CDM model, $67.4 \pm 0.5$  km s$^{-1}$ Mpc$^{-1}$ \citep{Planck2020}. Other tensions have also been revealed \citep[for the most recent review, see][]{eleonora2022}. On the other hand, some local measurements indicate much less tension, if any, when compared with the CMB measurements \citep[e.g.][]{freedman2021}. Therefore, systematic errors in the measurements could be an issue. In these circumstances, having more independent methods can help to firmly establish whether the tensions are genuine and whether we need new physics to explain the evolution of the Universe. These efforts would take us beyond the standard $\Lambda$CDM model.

Active Galactic Nuclei (AGN) are very complex systems; nevertheless, enormous progress has transpired in studying their nature since the early discovery days of the 1960s \citep{schmidt1963,salpeter1964,lyndenBell1969}. It is now well established that a typical AGN contains a supermassive black hole, which has been imaged, now for the two nearest sources -- M87 and Sgr A* -- with the Event Horizon Telescope \citep{EHT_M87_2019, EHT_Sgr_2022}. The intense radiation of AGN is powered by the accretion process. In bright AGN, the central black hole is surrounded by an optically thick, geometrically thin, and relatively cold accretion disk. Whereas in faint AGN, such as Sgr~A* or M87, the accretion flow is in the form of optically thin and much hotter plasma \citep[see][for recent reviews]{netzer2015,2021bhns.confE...1K}. The observed level of activity depends predominantly on the black hole mass, the accretion rate, and the viewing angle towards the nucleus as the AGN system possesses overall axial symmetry (but not spherical symmetry due to the presence of the disk), the collimated jet outflow perpendicular to the disk, and the circumnuclear dusty or molecular torus in the disk plane. These components are presumed to be present in highly accreting systems. 

Nowadays, broadband spectra of AGN cover all energy bands, from radio through the IR, optical, up to X-rays, and frequently to the gamma-ray band. Spectra are complex and full of atomic features - emission lines and absorption lines, atomic continua, and pseudocontinua. In addition, AGN are strongly variable in all wavelength bands. The cores of AGN are small and observationally unresolved, apart from extended radio and IR emission. But the variability and rich spectral features give us insight into the core structure using light echo studies done in the optical, X-ray, and IR  \citep[see][for the most recent review]{cackettKara2021}. AGN can be employed for cosmological distance measurements in many independent ways, but most methods rely on some assumptions about their structure.

\section{Inner boundary condition of the accretion disk surrounding a black hole}\label{sect:Inner}

The crucial step forward in studies of the accretion process onto a black hole happened with \citet{SS1973}, who found a simple and elegant way to parameterize the viscosity acting in a cold accretion disk. The original formulation of the model used the Newtonian description of the gravitational field of a black hole. However, at that time, the trajectories of the orbits in the Schwarzschild metric (non-rotating black hole) and in the equatorial plane of the Kerr metric (rotating black holes) were already known \citep{bardeen1972} and the pivotal role of - the innermost stable circular orbit (ISCO), was recognized. The existence of the ISCO is a characteristic property of general relativity (GR) that is absent in the Newtonian theory of gravity. Shakura \& Sunyaev imposed ISCO position as the innermost radius of the disk where the viscous torque can be assumed to vanish. The flow description close to the ISCO and below it, is not incorporated in the \citet{SS1973} model. 

This limitation started a discussion about what is actually happening in the inner region and how the flow proceeds towards the black hole horizon \citep[e.g.][]{fishbone1976,paczynski1981,loska1982}, and the problem of the transition between the Keplerian disk above the ISCO and the flow through the ISCO towards the black hole horizon was solved in full by \citet{muchotrzeb1982} (note that Muchotrzeb is the maiden name of the first author of this paper). This solution required the introduction of key terms into the dynamical and energy equations: (i) the description of the gravitational field in the pseudo-Newtonian approximation (which predicts the existence of an ISCO like in the Schwarzschild metric); (ii) the radial pressure gradient that allows for the deviation from Keplerian flow when pressure suddenly drops; and (iii) the advection term. This formalism finally revealed the nature of the accretion flow: the outer disk is roughly Keplerian, as in \citet{SS1973}, with a relatively slow radial velocity; close to the ISCO the inflow accelerates. Soon after crossing the ISCO the flow changes from subsonic to supersonic, and the angular momentum becomes roughly constant, so not much dissipation occurs below the ISCO. The solution showed that the model of \citet{SS1973} describes the cold disk flow well, and the departure in the dissipation profile is not very large up to the accretion rate corresponding to the Eddington luminosity. 

The conclusion that the viscous torque is close to zero at ISCO in the case of thin disks was generally accepted, even if the paper was based on the viscosity parametrization introduced by \citet{SS1973}, and, since 1991, it is well established that the accretion disk viscosity is, in fact, caused by the magnetorotational instability \citep{balbus1991}. However, there may be exceptions from this general statement \citep[see][and subsequent references]{stoeger1976}. First, it was noticed that too large a value of the  viscosity parameter leads to transonic flow well above the ISCO. Hence, all numerical solutions are ambiguous after crossing the sonic radius since the topology of the solution changes \citep{muchotrzeb1983,matsumoto1984,muchotrzeb1986}. Also, \citet{ashfordi2003} acknowledge that if a strong magnetic field is present, the flow conditions at the ISCO can vary.

Standard cold accretion disks are not the only form of the flow onto black holes. At high accretion rates, a slim disk develops, which will be discussed in the next section. While at lower accretion rates, a hot, geometrically thick and optically thin flow develops \citep{ichimaru1977,narayan1994}. There, a strong magnetic field can be present, so two branches of solutions must be considered: Magnetically Arrested Disks (MADs), where the strong magnetic field modulates or stops the flow \citep{MAD2003}, and the Standard And Normal Evolution (SANE) solutions, where the torque close to the ISCO remains small. 

\section{Stability of the radiation-pressure dominated accretion disk}\label{sect:Rad_Pres}

Soon after the introduction of such an elegant, simple, and practical accretion disk model by \citet{SS1973}, somewhat paradoxically, it was shown that those disks are thermally \citep{pringle1973} and viscously \citep{lightman1974} unstable. These instabilities happens when radiation pressure dominates over gas pressure (see also \citealt{SS1976} for a combined study of these two instabilities). This scenario is not only expected at high accretion rates in the case of binary black holes but even at moderate accretion rates in AGN. It had a lasting effect on the use of the standard accretion disk in AGN theory since many researchers considered that the instability must lead to disk destruction. On the other hand, broad-band spectra of bright AGN clearly showed the presence of the Big Blue Bump that dominates in the optical/UV band, which was fitted using a standard disk, eventually with some modifications \citep[e.g.][]{malkan1983,czernyElvis1987}. More recent broad-band spectra are of such high quality and allow for a determination of all disk parameters, including the spin of the black hole, which affects the exact position of the ISCO \citep[e.g.][]{czernySpin2011,capellupo2015}. The issue of disk stability is still not resolved. But significant steps have been made in this direction.

\subsection{Slim disk theory}

Introduction of all terms necessary to describe the accretion disk close to the ISCO \citep{muchotrzeb1982} led in a straightforward way to an important insight into the disk stability although this took a few more years to realize. \citet{abramowicz1988} showed that when the accretion rate exceeds the critical rate (corresponding to the Eddington luminosity for spherical accretion) the advection term becomes important. And an increasing fraction of energy dissipated in the disk is carried inward under the black hole horizon instead of being reemitted locally by the disk. The accretion disk is geometrically thicker than the standard disk, although not very thick (hence the name {\sl slim disk}). The emitted total radiation flux saturates with the ever-increasing accretion rate, and the cooling mechanism stabilizes the disk; a new advection-dominated solution branch emerges. First models assumed the pseudo-Newtonian potential, but later 1+1-D models in the full GR formalism, were also constructed \citep{sadowski2011}. These newer models included: (i) a proper radial structure in the equatorial plane; and (ii) the whole complexity of the disk's vertical structure earlier calculated in Newtonian models.\footnote{e.g. \citealt{rozanska1999} that already included complex opacity tables including molecules and dust grains, vertical convection with elements of the mixing-length theory, etc.}

The introduction of the slim disk concept encouraged the use of accretion disk models in AGN studies since the fate of the disk under immense radiation pressure was clear. At a given radius, the disk was stable when the gas dominated - at higher accretion rates, it was unstable, but at still higher accretion rates, it became stable again, with the three solution branches covering the whole parameter range. Since the transition between the branches depends on the radius, local stability or instability does not imply a global behaviour, but they provide a direct model prediction. If the accretion rate is so low that radiation pressure is never important the disk is globally stable. On the other hand, if the accretion rate is higher, the disk model shows the clear time dependence of a limit-cycle character, with the disk structure oscillating between the lower, colder gas-dominated branch and the upper, hotter advection-dominated branch. Such models assuming either a standard viscosity law as in \citet{SS1973} or simple modifications of this law were studied by several groups \citep[e.g.][]{szuszkiewicz1998,honma1991,janiuk2002,grzedzielski2017}. 

\subsection{Models of the limit cycle and the comparison to the observational data}

The introduction of the slim disk concept showed that the worst that can happen to an unstable disk is an onset of limit-cycle behavior. However, the parametrization introduced by \citet{SS1973} was just a simple scaling idea. Here, the radiation pressure instability strongly depends on the form of adopted scaling (for example, scaling of the viscous torque with only the magnetic pressure in equipartition with gas switches off the instability independently of the level of radiation, \citealt{sakimoto1981}). The best test of how the effective scaling of the viscous torque works in various objects is to test the models against observational data and to compare the stability parameter range, the outburst amplitudes, and the timescales.  In the case of Galactic black hole binaries, tests are relatively simple since the timescale of the predicted limit cycle is very short (tens to a few thousands of seconds). Moreover, the models based on the radiation pressure instability well describe the heartbeat states in this case \citep[e.g.][]{janiuk2000,janiuk2002,janiuk2015,rawat2022}. When we go to larger black hole masses, the model offers an interesting possibility of explaining the 400-day period in one source (HLX-1), which most likely contains an intermediate-mass black hole \citep[e.g.][]{sun2016,wu2016}. The application of this variability mechanism to AGN is based primarily on statistical studies \citep[e.g.][]{czerny2009}, or radio images \citep[e.g.][]{schawinski2015}, as the expected timescale of the process was hundreds to thousands of years. The general pattern across the entire black hole mass range is well-illustrated by \citet{wu2016}. 

\begin{figure}
\centering
\includegraphics[width=0.495\textwidth]{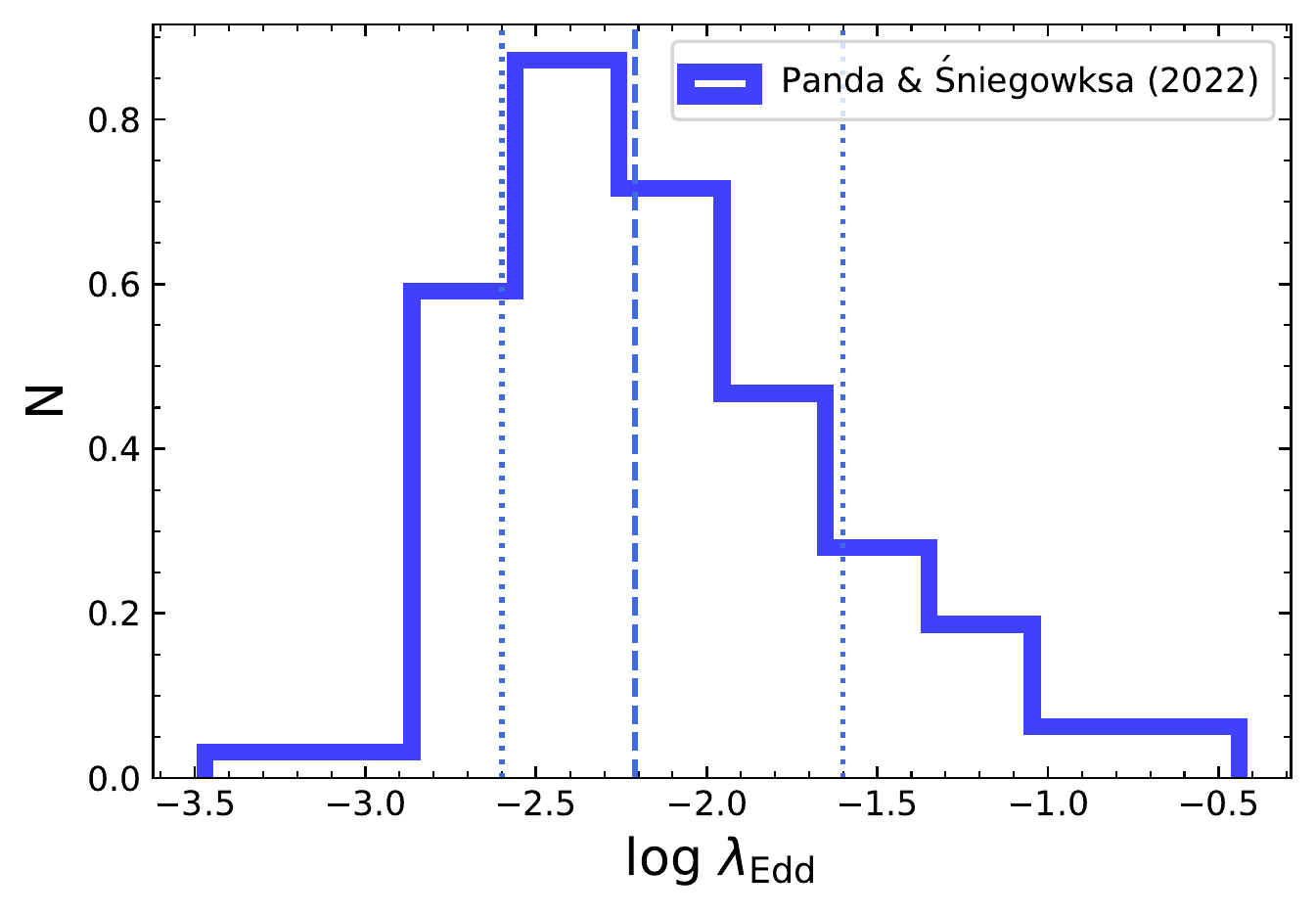}
\includegraphics[width=0.495\textwidth]{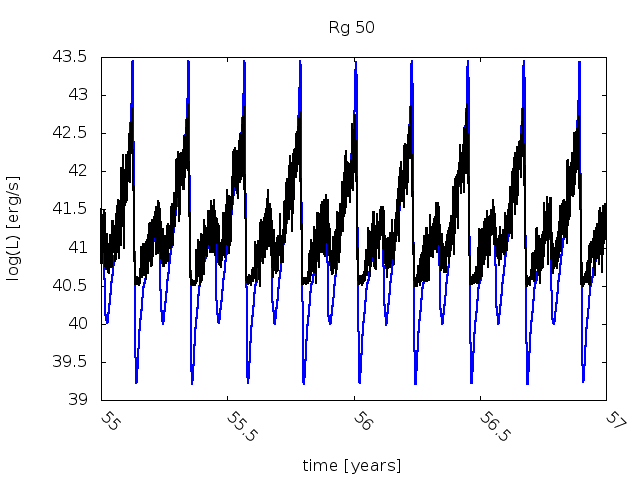}
\caption{\label{fig:GSN} Left panel: The distribution of the Eddington ratio in a sample of Changing-Look AGN listed by \citet{panda2022}, with the median value of $\log L/L_{\rm Edd} = -2.21$; other lines mark one-sigma error. Right panel: an exemplary model of the lightcurve from the accretion disk (blue line) and corona (black line) as inferred from GLADIS code computations for the case of black-hole mass $10^8 M_{\odot}$ and a small outer radius $50 R_g$. In this case, the characteristic timescale of the limit cycle caused by radiation pressure instability is short, of order of a month. The lightcurve is very complex.}
\end{figure}

Recently, there were several observations suggesting sudden changes in the innermost structure of some galactic nuclei, in the form of the appearance or the disappearance of broad emission lines \citep[e.g.][]{lopez2022,green2022}, a sudden change of X-ray properties \citep[][]{2003MNRAS.342..422M}, or the appearance of radio activity \citep[e.g.][]{Nyland2020}. The term ``Changing Look AGN'' was introduced \citep{2003MNRAS.342..422M} although it is unclear if a single mechanism is responsible for all these changes. These phenomenon may be partly caused by temporary obscuration \citep{goodrich1989} although recent studies seem to imply intrinsic variations in the sources \citep[e.g.][]{sheng2017} on the basis of IR observations. Another possibility is to have a tidal disruption of a star close to the supermassive black hole \citep[][]{zhang2021}. In some other sources, however, the phenomenon appears to be semi-regular on the timescales of a few years where the typical value of the Eddington ratio in these sources is of the order of 1 \% of the Eddington limit \citep[see Figure~\ref{fig:GSN}, left panel, and ][for the most recent compilation]{panda2022}. Multiple outbursts on a yearly timescale may be related to the relatively narrow radiation pressure instability zone \citep{sniegowska2020}, although more recent computations with the GLADIS code \citep{GLADIS2020J} indicate that in order to have such short timescales it is necessary to additionally reduce the outer radius of the disk to prevent the instability zone from spreading (see Figure~\ref{fig:GSN}, right panel, for an example). 

This could be the case if the disk was actually formed by (i) a tidal disruption event or (ii) if a secondary relatively massive black hole is present in the equatorial plane of the disk that could open a gap in the disk (\citealt{sniegowska2022}, also \v{S}tolc et al. 2023, submitted). The most extreme case of the semi-periodic activity of AGN are the Quasi-Periodic Eruptions (QPEs) observed in a growing number of sources \citep[e.g.][]{miniutti2019,chacraborty2021,2022MNRAS.515.4344K}. Modeling them with radiation pressure instability requires extreme parameters \citep{sniegowska2022}, so a TDE of a very compact star like a white dwarf is a very attractive alternative \citep{miniutti2022}. In principle, a recurrent star-disk or a compact remnant-disk interaction can also address some of the QPE properties. In addition, 2D and 3D GRMHD models imply the quasi-regular phenomena associated with nuclear outflows and jets \citep{2021ApJ...917...43S}. 

\subsection{Numerical MHD simulations and the soft X-ray corona}

An alternative way to establish if the radiation pressure instability develops in accretion disks and leads to the outbursts is to model the time-dependent behavior of the disk numerically. This can be achieved through magneto-hydrodynamic (MHD) simulations which account for the viscous torque acting in the disk without referring to {\em ad hoc} viscosity prescriptions. However, such simulations are very complex and computationally demanding. Computations performed in a shearing-box approximation did not show the instability \citep{Turner2004}, and first global computations confirmed that conclusion \citep{Hirose2009}. In contrast, later simulations with a denser grid exhibited the presence of instability \citep{Jiang2013}: the effect is sensitive to the details of the simulations, including the cooling description (and opacity), the initial magnetic field, the grid, and the duration of the computations. Most recent calculations suggest large variations in the predominantly thermal timescales \citep{Jiang2020}.

The radiation pressure instability may actually be damped when the magnetic field is substantial. There is research that suggests that instead of entering the limit cycle, the disk becomes strongly vertically stratified, with an inner, cold disk and an outer warm, optically thick and magnetically dominated corona \citep[e.g.][]{czerny2003,kubota2018,rozanska2015}. This model is an attractive explanation of the presence of the soft X-ray excess in most bright AGN \citep[see e.g.,][and references therein]{pop2020}. Thus, the issue of the radiation pressure instability in accretion disks is yet, in completeness, to be solved.

\section{Broad Line Region}
\label{sect:BLR}

Unobscured high-luminosity AGN show strong and broad Balmer lines \citep{schmidt1963}. Such emission lines are absent, or lines are double-peaked in Low Luminosity AGN where the inner part of the flow is in the form of hot advection-dominated accretion flow \citep[ADAF; see e.g.][and references therein]{Li2016}. These broad emission features come from clouds orbiting the black hole at a few hundred gravitational radii.\footnote{where the gravitational radius $R_{\rm g} = GM/c^2$, and $G$ is the gravitational constant, $M$ is the black hole mass, and $c$ is the speed of light} In addition, this region has extended structure and is stratified, as inferred from light echo studies that indicate a range of time delays between the continuum and the various emission lines in a given object \citep{wandel1999}. The fact that broad lines also exist when a colder disk is present strongly suggests a link between cloud formation and the standard disk itself \citep[e.g.][]{czerny2004}. 

The broad lines are, phenomenologically, divided into two populations: High Ionization Lines (HIL) coming from hotter lower-density clouds and Low Ionization Lines (LIL) originating from cooler and denser mediums. Measured time delays are shorter for HIL like He II and C IV, and longer for LIL represented by Balmer lines and Mg II emission\citep{peterson1999, 2019FrASS...6...75P, Panda_CaFe2021, zajacek2021, Panda2022_FrASS}. 

An attractive way of explaining the origin of the irradiated material was proposed by \citet{murray1995} who, developed a model of the line-driven wind from accretion disks. This model is a very efficient driving mechanism and a natural source of HIL. Later numerical studies supported the wind launching mechanism \citep{proga2000}, and recent, more advanced simulations address the issue of cloud formation due to thermal instability in such a wind \citep{waters2022}. 

\begin{figure}
\centering
\includegraphics[width=0.495\textwidth]{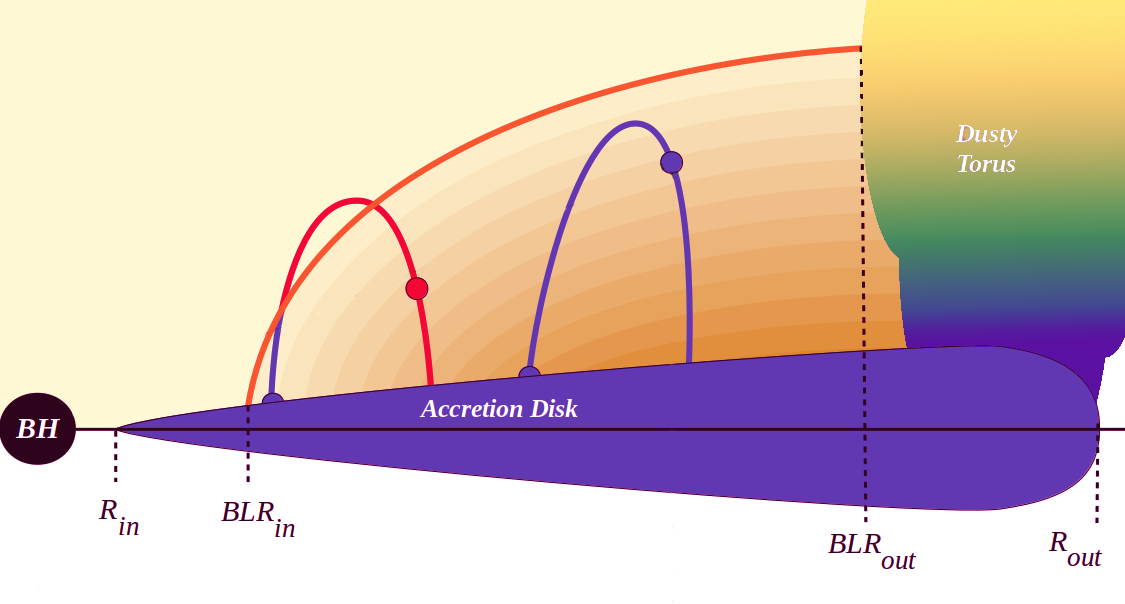}
\includegraphics[width=0.495\textwidth]{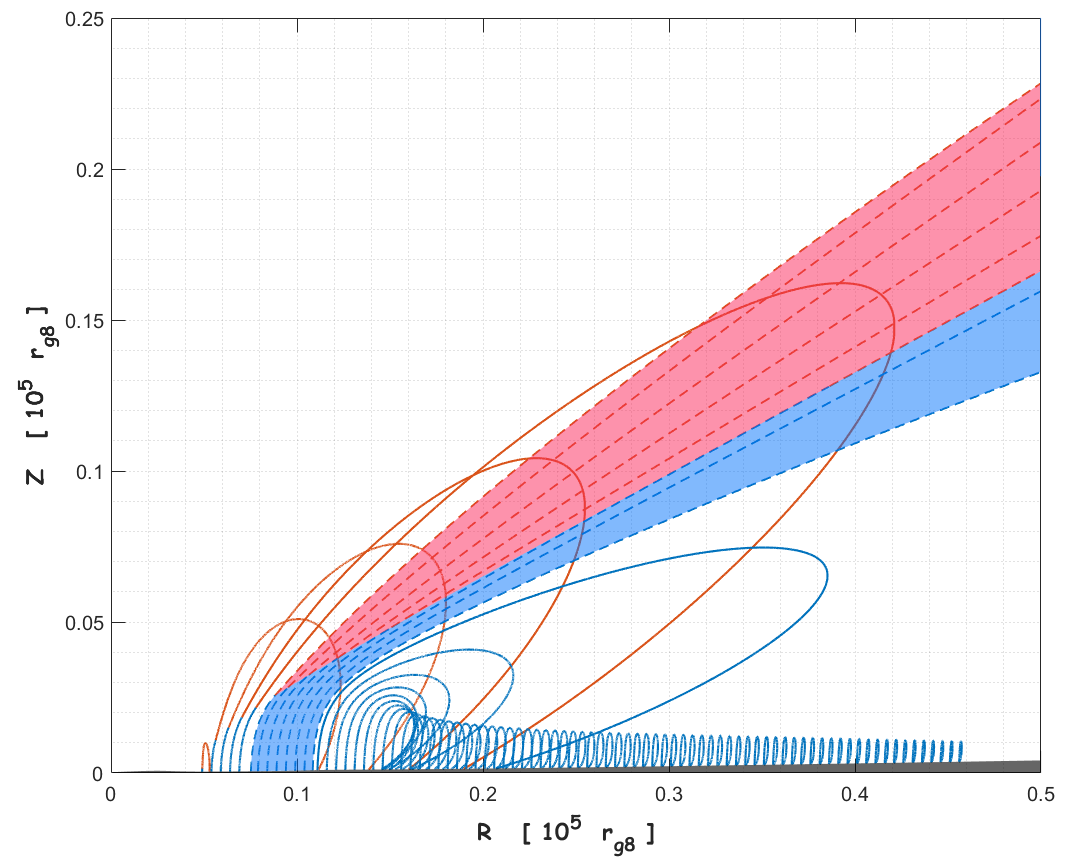}
\caption{\label{fig:FRADO} Left panel: The sketch of the FRADO model of the dynamics in the BLR. Dusty clouds are lifted and then fall back onto the disk due to the increase in the gravitational vertical component with height or, in addition, due to dust evaporation and the loss of radiation-pressure support. The blue line marks the trajectory of a dusty cloud, whereas the red line represents a dustless cloud in a ballistic motion. Right panel: Actual cloud trajectories for a black hole mass of $10^8\, M_{\odot}$, Eddington ratio of 1, and metallicity equal to 5 times the Solar value. A narrow stream of escaping clouds forms, with the red-shaded part denoting a dustless stream, while the blue-shaded part is dusty, and the failed wind develops in the outer part. All clouds are initially in Keplerian motion, and later on, angular momentum is preserved during the 3-D orbital motion.}
\end{figure}

In stars, the line-driven wind is not the only mechanism, and most massive winds come from stars with dusty envelopes like AGB stars. Therefore \citet{czhr2011} proposed a dust-driven wind as a source of material for LIL in AGN. The effective temperature of the accretion disk decreases with the radius, and at some distance, it drops below the dust sublimation temperature, which is of order 1000 K. This is not yet the region that is traditionally identified with the dusty torus since the AGN disk is not strongly irradiated due to its flat geometry \citep{loska2004}. The dusty torus distance is estimated for efficient irradiation by the central regions. However, in the \citet{czhr2011} model, when the material is lifted high above the disk, it becomes irradiated, dust evaporates, and falls ballistically back towards the disk, thus forming the FRADO - Failed Radiatively Accelerated Dusty Outflow. We show the sketch of the model in Figure~\ref{fig:FRADO}, left panel. The onset of the dusty wind is uniquely determined by the sublimation temperature. This distance from the black hole does not depend solely on the black hole mass and accretion rate, but on a combination of these two parameters set by the absolute monochromatic luminosity. This is one of the great advantages of this model. 

Numerical 2.5-D computations of cloud motion under dust radiation pressure show that the cloud motion is complex and depends strongly on parameters like the black hole mass, the accretion rate, the viewing angle towards the nucleus, and the metallicity of the material \citep{naddaf2021,ana2022,naddaf2022}. At low accretion rates, the wind has the character of a failed wind, with a relatively small vertical velocity component on top of the azimuthal motion. But, at high accretion rates and/or high metallicities at some range of radii, an escaping wind forms with a vertical velocity comparable to the azimuthal component (see Figure~\ref{fig:FRADO}, right panel). Such an escaping wind forms a structure similar to the one proposed by \citet{elvis2000}. A similar geometrical structure was also recently put forward by \citet{hoenig2019} for the IR and sub-mm band, although this structure is located at a larger distance from the black hole by a factor of 10--100 as compared to the region implied by the FRADO.

The stability of the BLR clouds within the FRADO model was recently investigated by \citet{ana2022} (see their Appendix), who also pointed to the potential prolongation of the clouds due to tidal forces acting on an orbital timescale. Despite the fast development of Kelvin-Helmholtz instability, cloud ablation, evaporation, and tidal stretching within one orbital timescale, thermal instability acting on the two-phase hot-cold medium \citep{1981ApJ...249..422K} is expected to lead to the condensation of the hot phase onto the colder cloud core. This way, the cloud is pressure-confined by the surrounding hot medium and can survive for a sufficiently long time so that it eventually collides with the disk as it falls back.

It remains to be seen if the HIL, LIL, and torus outflows form separated nested structures. This scenario could be tested further by light echo studies and computational models which combine line driving and dust driving mechanisms. Launching an outflow at such large distances as required by the \citet{hoenig2019} model will additionally require modification of the structure of the outer disk, which is possibly related to gravitational instability \citep[see e.g.][]{thompson2005,czerny2016}.

\section{Cosmology with quasars}

A general idea that quasars should be suitable for cosmology to test the properties of the Universe was already published in 1973 by Setti \& Woltjer in their paper titled {\it Hubble Diagram for Quasars} \citep{woltier1973}. However, at that time, it was far too early for an actual practical use of the proposed method, based on just plotting $V$-magnitude of radio-loud quasars against redshift $z$. With progress in the understanding of AGN structure, and numerous observations of a different character, several methods were proposed to use quasars for distance measurements and tracing the Universe expansion. Most of these methods are discussed in the review by \citet{czerny_SSRv2018}. In this review, we will concentrate on the methods based on light-echo studies. These methods can be divided into two main groups: (i) accretion disk continuum delay mapping, and (ii) broad emission line time delays with respect to the continuum.

\begin{figure}
\centering
\includegraphics[width=0.495\textwidth]{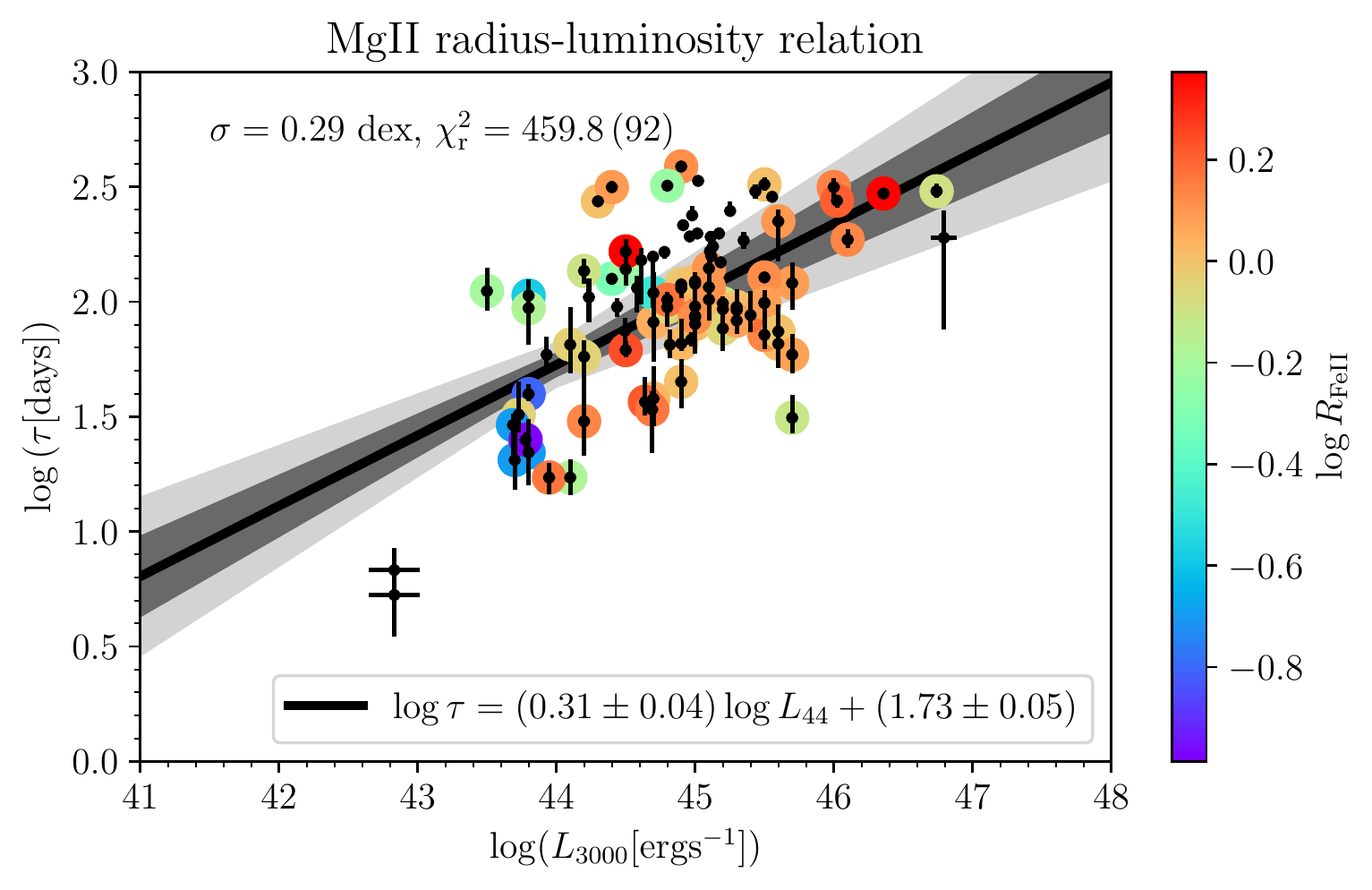}
\includegraphics[width=0.495\textwidth]{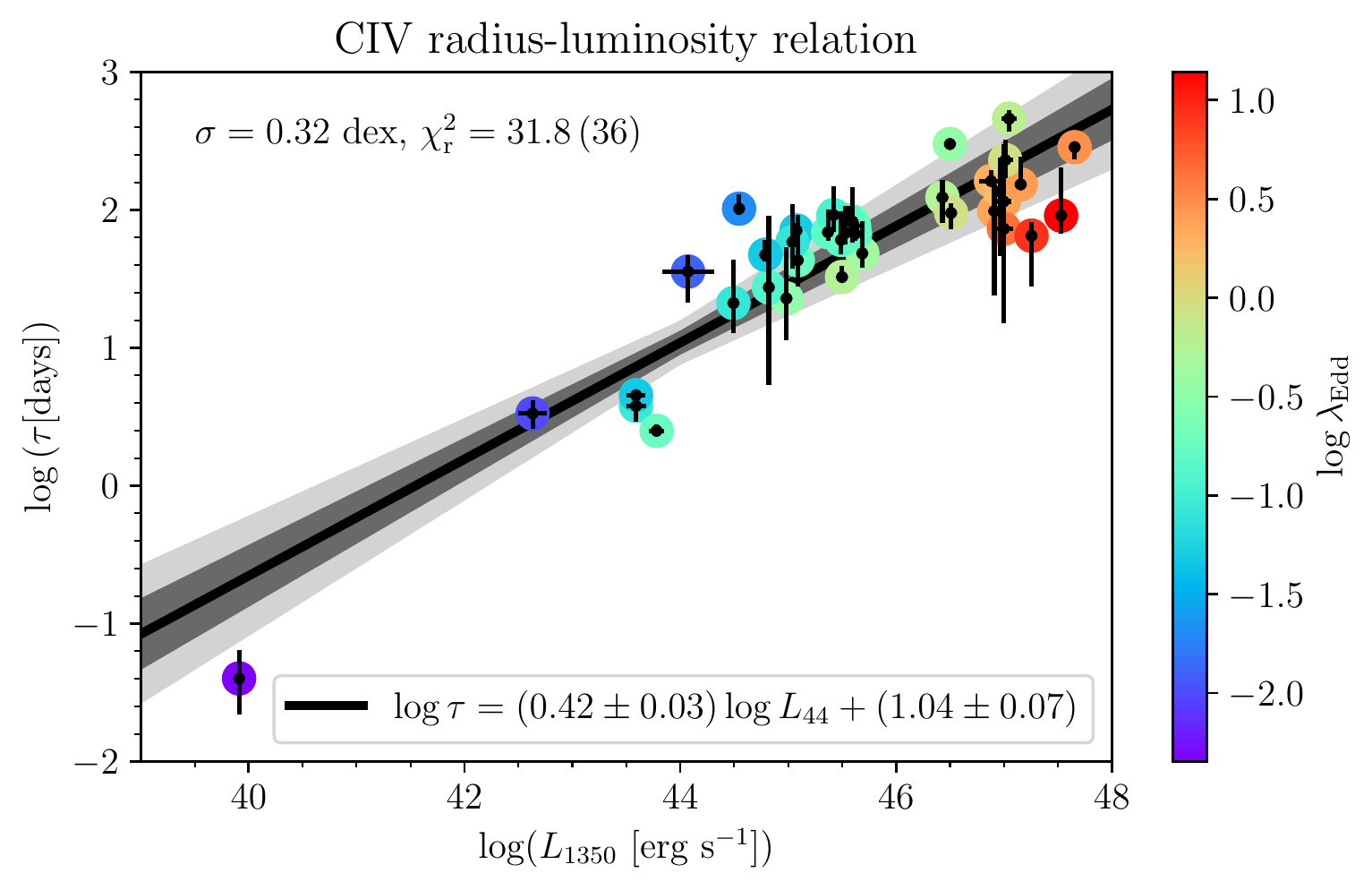}
\caption{\label{fig:R_L} The radius-luminosity relation for the Mg II (left panel, 94 sources, see \citealt{khadka2021} and \citealt{2022arXiv220805491Y}) and C IV (right panel, 38 sources, see \citealt{2021ApJ...915..129K} and \citealt{cao2022}) samples in the spatially flat $\Lambda$CDM model ($H_0=70\,{\rm km\,s^{-1}\,Mpc^{-1}}$, $\Omega_{\rm m0}=0.3$). For the Mg II radius-luminosity relation, the sources are color-coded by the relative Fe II strength ($\log{R_{\rm FeII}}$), i.e.\ for those which the measurement is available (66 sources). For the C IV radius-luminosity relation, sources are colour-coded according to the estimated Eddington ratio ($\log{\lambda_{\rm Edd}}$). The shaded grey areas around the best fit lines stand for 1$\sigma$ and 2$\sigma$ confidence intervals. The best-fit radius-luminosity relation parameters, the vertical RMS scatter and the reduced $\chi^2$ are given in each panel.}
\end{figure}

At present, the second method is more advanced, although it was proposed only in 2011 \citep{watson2011,Haas2011}. In this method, one uses the general linear parametrization between the radius of the BLR and the monochromatic absolute luminosity. This radius - luminosity relation has two free parameters - the proportionality constant and the intecept. Such correlations are established for a few strong emission lines. Two examples of such trends, for Mg II and C IV sources, are shown in Figure~\ref{fig:R_L}, together with the best linear fit. In principle, models like FRADO (discussed in Section~\ref{sect:BLR}) allow us to avoid this parametrization. Although the region is extended, and not easily replaced with the effective radius, making the parametric approach safer. Cosmological constraints derived based on Mg II and C IV emission lines are basically consistent (so far) with the standard $\Lambda$CDM model and do not imply any tension \citep{zajacek2021,khadka2021,cao2022}. The Fe II strength has been shown in previous studies to reproduce the accretion rate effect - the shorter-than-expected lags at higher luminosities is proportional to the accretion rate \citep{Du_Wang_2019, Panda2022_FrASS}. But, taking into account the additional dependence of the time delay on the Fe II strength does not improve the cosmological constraints \citep{2022MNRAS.515.3729K}.

\begin{figure}
\centering
\includegraphics[width=0.975\textwidth]{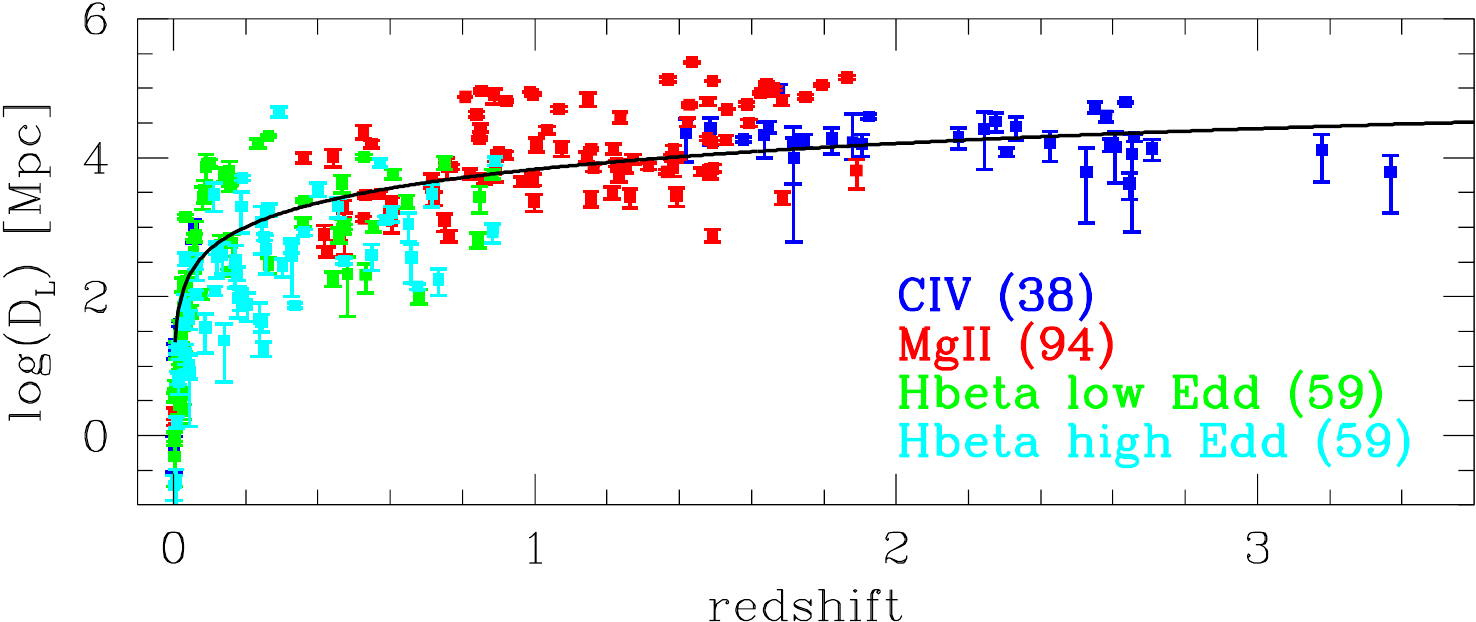}
\caption{\label{fig:Hubble} The location of quasars studied by \citet{khadka2021},  \citet{cao2022} and \citet{khadka2022} on the Hubble diagram, with H$\beta$ sources divided into high and low Eddington ration subsamples.  Numbers in parentheses show the number of measurements. The black line shows the flat $\Lambda$CDM cosmological model favored by the common fit of AGN (Mg II, C IV), chronometry and baryon acoustic oscillations \citep{cao2022}, with cosmological parameter values $H_0 = 68.86$ km s$^{-1}$ Mpc$^{-1}$ and $\Omega_m$ = 0.295.}
\end{figure}

The use of the H$\beta$ line for this purpose is also possible in principle but faces more problems - either related to the fact that time-delay measurements are collected from various groups using different time-delay techniques or due to the strong dependence of the delay on the source accretion rate \citep[e.g.][]{maryloli2019,khadka2021}. The location on the Hubble diagram of all the quasars studied by us is shown in Figure~\ref{fig:Hubble}. We can see that the scatter seems larger for the H$\beta$ subsample. The number of AGN with available time-delay measurements is of order a 100, and this can also contribute to relatively weak constraints provided so far by quasars in the context of this method. 

These radius-luminosity relations, for various emission lines, can also be combined with other independent local tracers of the Universe expansion. For example, the Hubble constant determined from chronometric measurements of the Universe expansion, baryon acoustic oscillation data, lower-redshift Type Ia supernovae, Mg II reverberation-measured quasars, quasar angular sizes, H II starburst galaxies, and Amati-correlated gamma-ray burst data, gives the value of $69.7\pm1.2$ km s$^{-1}$ Mpc$^{-1}$. This value is consistent with the Planck CMB data value \citep{2022MNRAS.513.5686C}.

\begin{figure}
\centering
\includegraphics[width=0.495\textwidth]{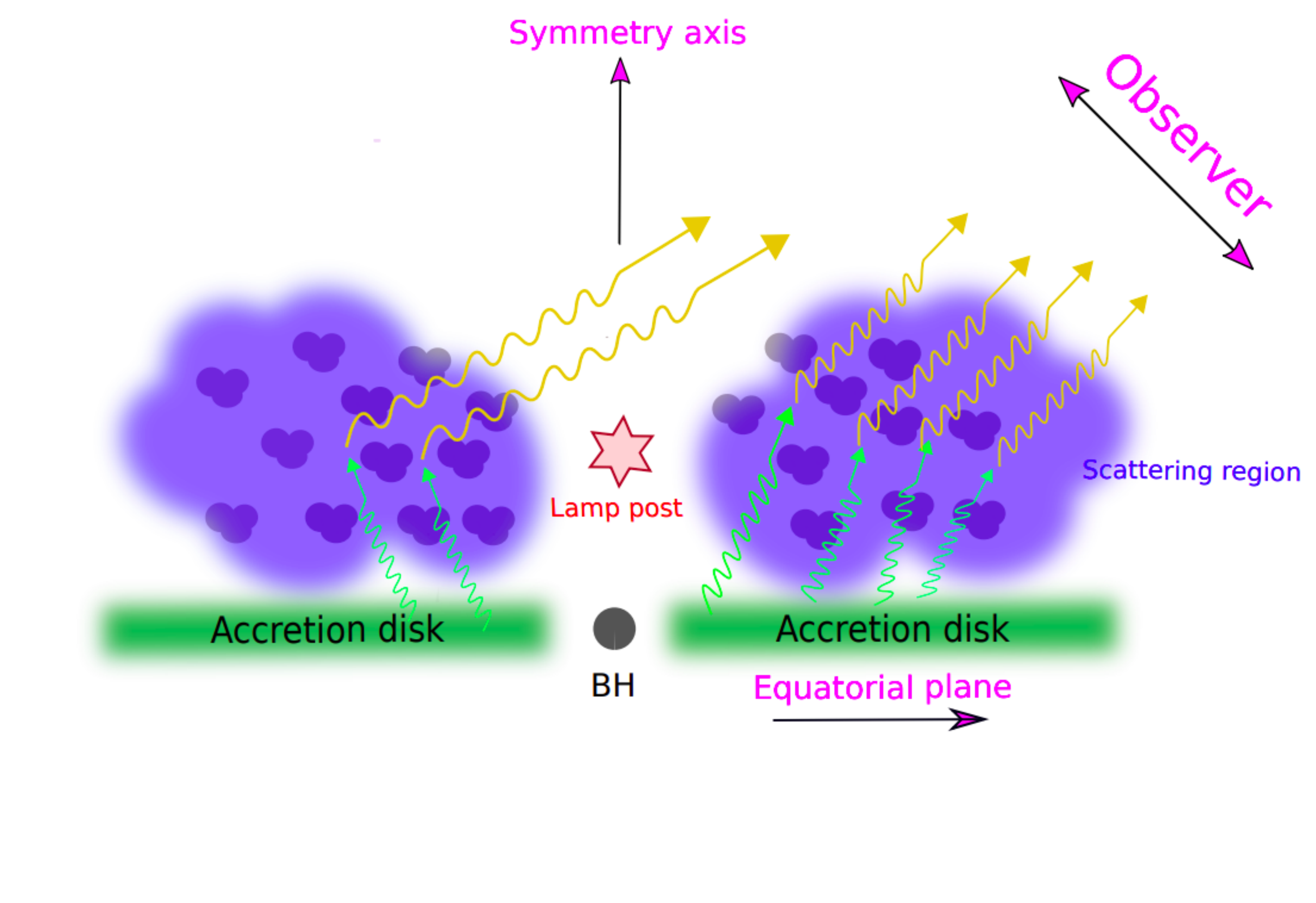}
\includegraphics[width=0.495\textwidth]{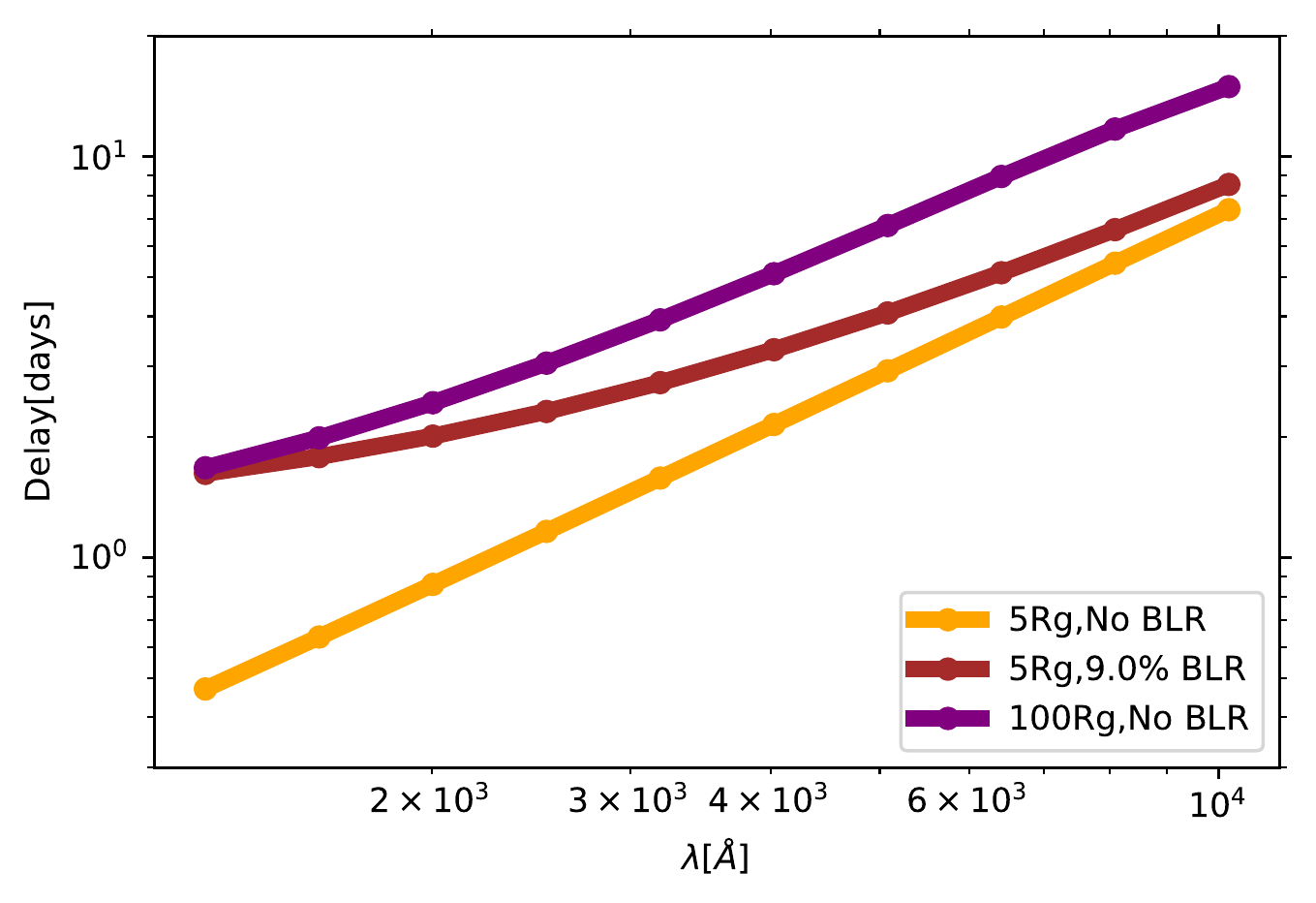}
\label{fig:vikram}
\caption{Left panel: Schematic picture of disk illumination by the corona and the subsequent scattering of disk photons in the inter-cloud medium. Right panel: The simulated time delay for a lamp-post model, with the source located at a distance of 5 $R_{\rm g}$ or 100 $R_{\rm g}$ with no BLR scattering, and with the effect of scattering of 30\% of photons by the BLR included. The shape of the delay curve is clearly different. Other model parameters considered in calculations are black hole mass of $10^8\, M_{\odot}$, Eddington ratio of 1, inclination angle of 30 degrees, and the luminosity of the corona is $3.78 \times 10^{46}$ erg s$^{-1}$. The BLR response shape is a half Gaussian function with peak at 5 days and a width of 10 days.}
\end{figure}

The first light-echo-based method was proposed back in 1999 \citep{collier1999}, but the applications towards Cosmology were not successful since the derived Hubble constant was far too low \citep{cackett2007}. This is generally known as the `too large disk size issue', identified independently through light echo and microlensing studies. Recent dense monitorings of a number of sources are beginning to indicate that the problem might be related to the additional reprocessing of the disk photons by the surrounding plasma, including the BLR itself that contaminates the signal by scattering (see Figure~\ref{fig:vikram}, left panel), absorption, and subsequent re-emission of the photons in the form of broad emission lines as well as Balmer continuum \citep{chelouche2019,netzer2022}. The estimated time delay also depends on the position of the irradiating source, the accretion disk albedo and the color correction to the disk effective temperature \citep{kammoun2021}. 

One way to distinguish between the effect of the position of the irradiating source and the contamination by the scattering medium (see Figure~\ref{fig:vikram}) is discussed in \citet{Jaiswal2022J}. However, in some sources, the effect may also be related to an incorrect estimate of the black hole mass \citep{pozo2014,pozo_509_2019}.

\section{Future prospects}

The amount of observational data will continue to increase in coming years. Several large, ground-based optical surveys are now in operation (e.g.\ Zwicky Transient Facility, ZTF, \citealt{Zwicky-TF_2019}), and others will start operating in the near future (e.g.\ Vera Rubin Observatory and its Legacy Survey of Space and Time -- LSST, \citealt{LSST_2019ApJ...873..111I,2019FrASS...6...75P,2022arXiv220806203K}). Some of these surveys have the general character of all-sky surveys, while some are focused on searching for transient sources or even specifically on reverberation mapping of AGN (e.g. Sloan Digital Sky Survey Reverberation Mapping - SDSS-RM, \citealt{2020ApJ...901...55H}). These surveys will increase the number of known AGN by ten times and the number of measured time delays by a factor of a hundred or more. Therefore, statistical errors in all measurements will decrease, which is particularly important for cosmology and the tension issue. However, \texttt{accuracy} and \texttt{precision} do not have the same meaning, and the issue of possible systematic effects will become increasingly important. 

Therefore, massive studies through extensive surveys are necessary, but there is also room for dedicated campaigns on selected sources, particularly for reverberation mapping of AGN. Valuable campaigns are underway using dense and multi-band monitoring as they shed light on the complexity of radiation reprocessing in AGN \citep[e.g.][]{edelson2015,edelson2019,kriss2019,vincentelli2022}, and similar studies are expected in the future, e.g. by using smaller UV telescopes \citep{2022arXiv220705485W}. Also, in massive surveys, extreme sources -- very faint or very bright -- are under-represented, and play a crucial role in determining trends like radius-luminosity relations. Nearby, faint sources require very dense monitoring, while bright distant quasars require long monitoring of several years to measure the delay of the broad emission lines \citep[e.g.][]{lira2018,2019ApJ...880...46C,2020ApJ...896..146Z,zajacek2021,2022A&A...667A..42P}. The time delay for the continuum has not been measured so far for distant quasars and can be done best with dedicated telescopes. Such dedicated, source-oriented monitoring combined with extensive simulations can also provide correction factors for massive surveys. We plan to carry out such monitoring effort at the Observatory Cerro Armazones (OCA) as a part of the program planned in our ERC project.


\bmhead{Acknowledgments}
This project has received funding from the European Research Council (ERC) under the European Union’s Horizon 2020 research and innovation program (grant agreement No. [951549]). The project was partially supported by the Polish Funding Agency National Science Centre, project 2017/26/A/ST9/00756 (MAESTRO 9), and MNiSW grant DIR/WK/2018/12. MLM-A acknowledges financial support from Millenium Nucleus NCN$19\_058$ (TITANs). VK acknowledges the Czech Science Foundation project No. 21-06825X ``Accreting black holes in the new era of X-ray polarimetry missions''. SP acknowledges financial support from the Conselho Nacional de Desenvolvimento Cient\'{\i}fico e Tecnol\'ogico (CNPq) Fellowship (164753/2020-6). F.\ Pozo Nu\~nez gratefully acknowledges the generous and invaluable support of the Klaus Tschira Foundation. MZ acknowledges the financial support of the GA\v{C}R EXPRO grant No. 21-13491X ``Exploring the Hot Universe and Understanding Cosmic Feedback''. SC, NK, and BR acknowledge US DOE grant DE-SC0011840. ZY is supported by Chandra grant GO9-20084X. NK would like to acknowledge Dr. Richard Jelsma (a Bellarmine University donor).
The authors also acknowledge the Czech-Polish mobility program (M\v{S}MT 8J20PL037 and
NAWA PPN/BCZ/2019/1/00069) and the OPUS-LAP/GA\v{C}R-LA bilateral project (UMO-2021/43/I/ST9/01352 and GF22-04053L). 

\section*{Declarations}

\begin{itemize}
\item Funding

The funding details are listed in Acknowledgement section.

\item Conflict of interest/Competing interests 

The authors declare they have no financial interests.

\item Ethics approval 

Not applicable

\item Consent to participate

All authors contributed to the work and approved sending the paper for publication.

\item Consent for publication

All authors agree for the work to be published in Astrophysics and Space Science.

\item Availability of data and materials

Data can be available upon request.

\item Code availability 

Codes which were made public (GLADIS) are provided with the reference and a link for downloads via Astrophysics Source Code Library.

\item Authors' contributions

All authors contributed to the study conception and design. B.C. was the leading author in writing the text.  S.C. and Z.Y. provided the data and helped with Fig. 4.  V.K.J. and R.P. prepared Fig. 5., V.K.J. helped considerably with text formating as well. M.L.M.-A. suggested Fig. 4 and provided part of the data.  All authors reviewed the manuscript. M.S. prepared Fig. 1, right panel. M.H.N. made Fig. 2. S.P. made Fig. 1, left panel.  V.K.,  F.P.N. and B.R. contributed significantly to the text. All authors commented on the text at the draft stage. Finally, all authors carefully reviewed and approved the final version of the text.
\end{itemize}







\bibliography{sn-bibliography}

\end{document}